\documentclass{article}
\usepackage{graphicx} 
\usepackage[utf8]{inputenc}
\usepackage{float}
\usepackage{amsmath}
\usepackage{setspace}
\usepackage{upgreek}
\usepackage{siunitx} 
\usepackage{graphicx}
\usepackage{amsmath, amssymb}
\usepackage{authblk}
\usepackage{url}
\usepackage{caption}
\usepackage[margin=3.2cm]{ geometry }
\usepackage{titlesec} 

\titleformat{\section}
  {\normalfont\large\bfseries}{\thesection}{1em}{} 
\title{\Large{Optical-Theorem-Based Holography For Target Detection and Tracking}}

\date{}
\author[1] {Mohammadrasoul Taghavi}
\author[1*]{Edwin A. Marengo}
\affil[1]  {Department of Electrical and Computer Engineering, Northeastern University, Boston, Massachusetts 02115, USA}

\begin{document}
\maketitle
    
\noindent{\bf{ Abstract}}
\vspace{0.4cm}

The development of robust, real-time optical methods for the detection and tracking of particles in complex multiple scattering media is a problem of practical importance in a number of fields, including environmental monitoring, air quality assessment, and homeland security. In this paper we develop a holographic, optical-theorem-based method for the detection of particles embedded in complex environments where wavefronts undergo strong multiple scattering. The proposed methodology is adaptive, to the complex medium, which is integral to the sensing apparatus, and thereby enables constant monitoring, through progressive adaptation. This feature, along with the holographic nature of the developed approach, also renders as a by-product real-time imaging capabilities for the continuous tracking of particles traversing the region under surveillance. In addition, the proposed methodology also enables the development of customized sensors that leverage a controllable complex multiple scattering medium and the derived holographic sensing technology for real-time particle detection and tracking. 
We demonstrate, with the help of realistic computer simulations, holographic techniques capable of detecting and tracking small particles under such conditions and analyze the role of multiple scattering in enhancing the detection performance. 
Potential applications include the identification of aerosolized biological substances, which is critical for biosecurity and the rapid detection of hazardous airborne particles in confined or densely populated areas.

\section{Introduction}
The ability to accurately detect and characterize particles is crucial in a wide range of scientific and practical applications related to environmental monitoring, air quality assessment, and national security \cite{simonet2009monitoring,badireddy2012detection,schmid2013real}. In this context, optical-sensing methods are popular as they enable real-time and in-situ examination of the nanoparticles present in typical media such as aerosols \cite{ignatovich2007optical,mudanyali2013wide, berg2022tutorial, farazi2024optical}. The present work is concerned with the detection and tracking of particles via holographic optical-theorem-based methods applied to complex scattering media. Two applicational contexts are considered. First, in many practical scenarios the sought-after particle detection and tracking needs to be implemented in conditions where the surrounding environment is composed of many multiple scattering constituents. The complex medium scrambles the wavefronts, yielding complex waves. This makes detection and imaging quite challenging in such media. On the other hand, the same complex multiple interactions also enhance the degree of interactivity with the target. This may potentially enhance, in principle, the detectability of small particles relative to simpler media, e.g., under free space or weakly interacting environments. Thus, the first objective is to demonstrate holographic methods that are feasible for the detection and tracking of small particles in such media. In addition, we also show ways in which the complex medium does, indeed, enhance detectability. The second objective is to examine the possibility of developing sensors that are based on the use of a complex, multiple scattering medium, along with holographic sensing equipment, for the detection and tracking of particles. Several applications are envisioned. One of the major applications to consider is the identification of aerosolized biological substances that could serve as biological warfare agents, among naturally occurring harmless particles, in a real-time and non-invasive manner \cite{frohlich2016bioaerosols}. Optical manipulation techniques provide a powerful approach for controlling the movement and interactions of nanoparticles in various environments \cite{urban2014optical,sadafi2023optical}. In addition, it has been demonstrated that engineered Janus particles can be used to create asymmetric imaging smoke clouds, enabling target concealment on one side while maintaining clear visibility on the other, with particle scattering properties optimized through advanced computational techniques \cite{sadafi2024electrostatic,da2024asymmetric}. 
In these contexts, it is crucial to have an apparatus that is effective, as well as possibly reconfigurable (e.g., via the holographic aspect), for tasks such as detecting intrusion, tracking medium changes, and performing imaging. 

\begin{figure}[H]
    \centering
    \includegraphics[width=1\linewidth]{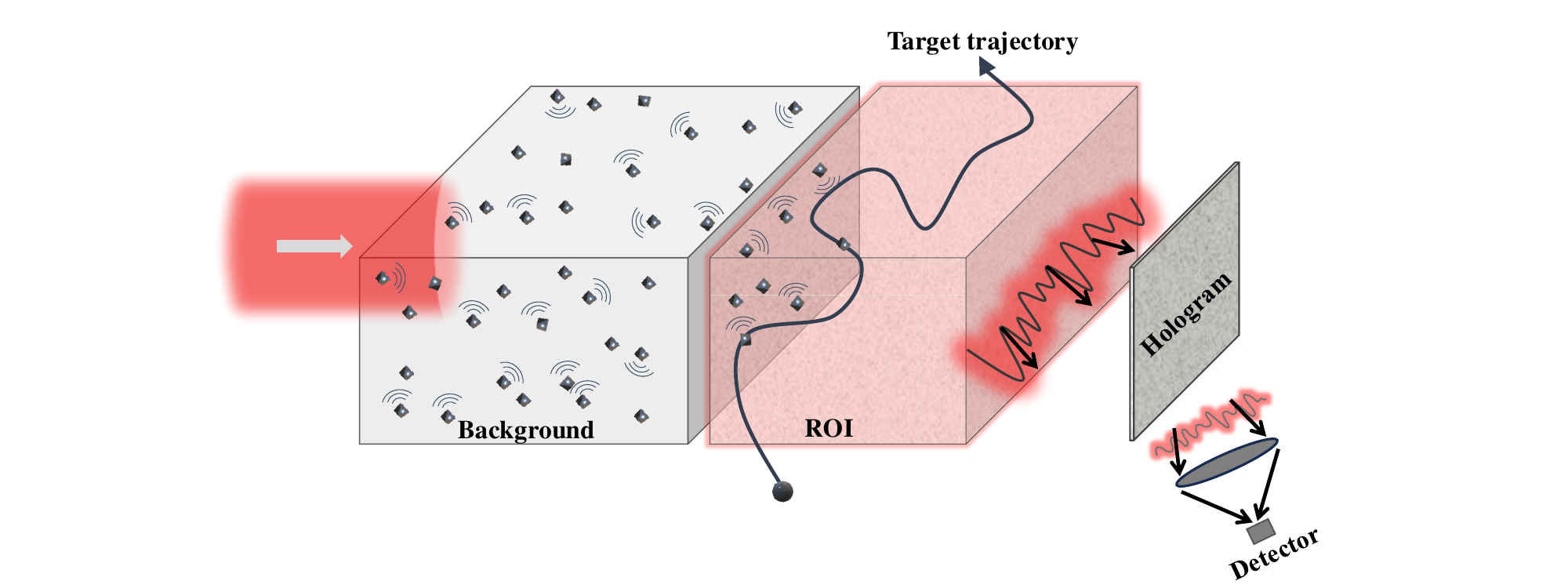}
    \caption{Conceptualization of the proposed holographic imaging system for the detection and tracking of targets traversing the ROI.}
    \label{fig0}
\end{figure}
The interaction between light and matter has become a prominent research topic in recent years \cite{babicheva2024mie,barati2024non, barati2023singular, farazi2018tunable, sadafi2023dynamic}. Furthermore, scattering behavior of nanoscatterers is highly sensitive to the structural deformation, and inhomogeneity of the refractive index of the particle cluster \cite{moon2023measuring,zhao2022speckle,taghavi2024dynamic}. Scattering-based techniques have been effective in determining key attributes such as particle shape, size and composition \cite{xu2001particle}. This makes scattering-based characterization of nanoparticle clusters such as aerosols potentially feasible. Thus in the present work we adopt scattering-based sensing methods. Figure~\ref{fig0} provides an schematic of the envisioned system. A laser beam is employed for the interrogation of a complex medium, which we assume to be composed of many randomly positioned scatterers with rather arbitrary scattering properties. The resulting perturbed waveform is adopted for probing of the region under surveillance or region of interest (ROI), as shown in the figure, for the purpose of detecting and tracking particle motions in that region. The gathered data is captured holographically at the output, rendering a single-pixel or bucket detector measurement that is representative of the extincted power or energy of the scattering phenomenon. This quantity is adopted for the purposes of particle detection while position-dependent variations of the associated response are adopted for the envisioned tracking. 

The sensing of the particle cross section in the complex medium relies on a fundamental principle in scattering theory, termed ``the optical theorem''. This is a  key principle in scattering theory, and it is quite universal. It applies to all the wave disciplines (electromagnetics, acoustics, optics, quantum mechanics, and so on). The optical theorem has wide applications in imaging, remote sensing, and target detection \cite{marengo2024optical, marengo2024optical2}. It was introduced by Wolfgang 
Von Sellmeier and Lord Rayleigh (John Strutt) in 1871 \cite{born2013principles, marengo2015nonuniqueness, marengo2015optical, marengo2013target}. Some recent studies have introduced methods for designing sensors that measure extincted power in diverse environments for arbitrary fields using time-reversal mirrors or cavities \cite{tu2016generalized, marengo2015nonuniqueness, marengo2016generalized,6710835}. However, this approach depends on phase information, making it less suitable for optical and quantum systems, where only field intensities are measured. On the other hand, the same principle can be employed in the optical regime through holographic techniques. Berg et al. has proposed and experimentally verified the possibility of measuring the extinction cross-section of spherical and non-spherical particles \cite{berg2017measuring, berg2014using} for plane-wave incidence. In this work we borrow from prior formulations of the optical theorem in complex media \cite{tu2016generalized, marengo2015nonuniqueness, marengo2016generalized,6710835} and combine them with the holographic sensing methods in \cite{memmolo2015recent}  for the development of a new optical theorem holography whose objective is the remote sensing of the extinction or cross section of the target. This fundamental measure is then utilized as an indicator of target presence, where large values of the measured cross section, over a suitably selected threshold, are indicative of target presence.

 Holography is vital in the optical regime, where it enables the recording of both amplitude and phase information associated to the information-carrying fields. Furthermore, holography has been widely utilized in scattering-based encryption systems as it offers advanced capabilities for secure information processing and imaging through complex media \cite{yu2022scattering,taghavi2024differential}. The proposed particle detection methodology relies heavily on holography, for the coherent processing of the scattered fields. There have been important prior related works in this area, including work on holographic methods for particle and biological sample tracking \cite{memmolo2015recent,su2012high,walcutt2020assessment,kim2024digital,potter2024clinical,elius2022effect,rogalski2022accurate, marengo2016holographic}. Holography has also been utilized in nanoparticle characterization applications to reconstruct silhouette-like images of the particle from interference pattern measurements \cite{berg2011digital}. Additionally, for detection and imaging of tiny scatterers, quantitative phase imaging using holography offers precise, label-free visualization, making it ideal for analyzing weakly absorbing samples and understanding their structural and optical properties \cite{huang2024quantitative,yang2018quantitative,xie2024quadri}. Ravasio et al. has utilized the digital holographic approach to simultaneously measure the optical extinction cross-section and morphological properties such as cross-sectional shape and size of single mineral dust particles \cite{ravasio2021optical}. In another work, Berg et al. introduced a simplified bi-telecentric lens system for digital in-line holography, enabling high-resolution imaging of free-flowing aerosol particles at the sub-micrometer scale \cite{berg2024imaging}. Also, holographic particle image velocimetry (HPIV) has emerged as a powerful technique for obtaining three-dimensional, time-resolved velocity field measurements in complex fluid flows by leveraging the intrinsic volumetric imaging capability of holography \cite{meng2004holographic,hong2025review}.

The remainder of the paper is organized as follows. In section 2, we outline the envisioned operational concept and proposed methodology for optical-theorem-based holographic detection and tracking of small particles in complex media. Section 3 provides computer simulations that demonstrate the proposed technique's effectiveness and advantages relative to alternative methods such as non-holographic detection methods. The results in Section 3 are based on rigorous computations including all the relevant multiple scattering interactions in the complex medium. Section 4 provides concluding remarks. 

\begin{figure}[H]
    \centering
    \includegraphics[width=1\linewidth]{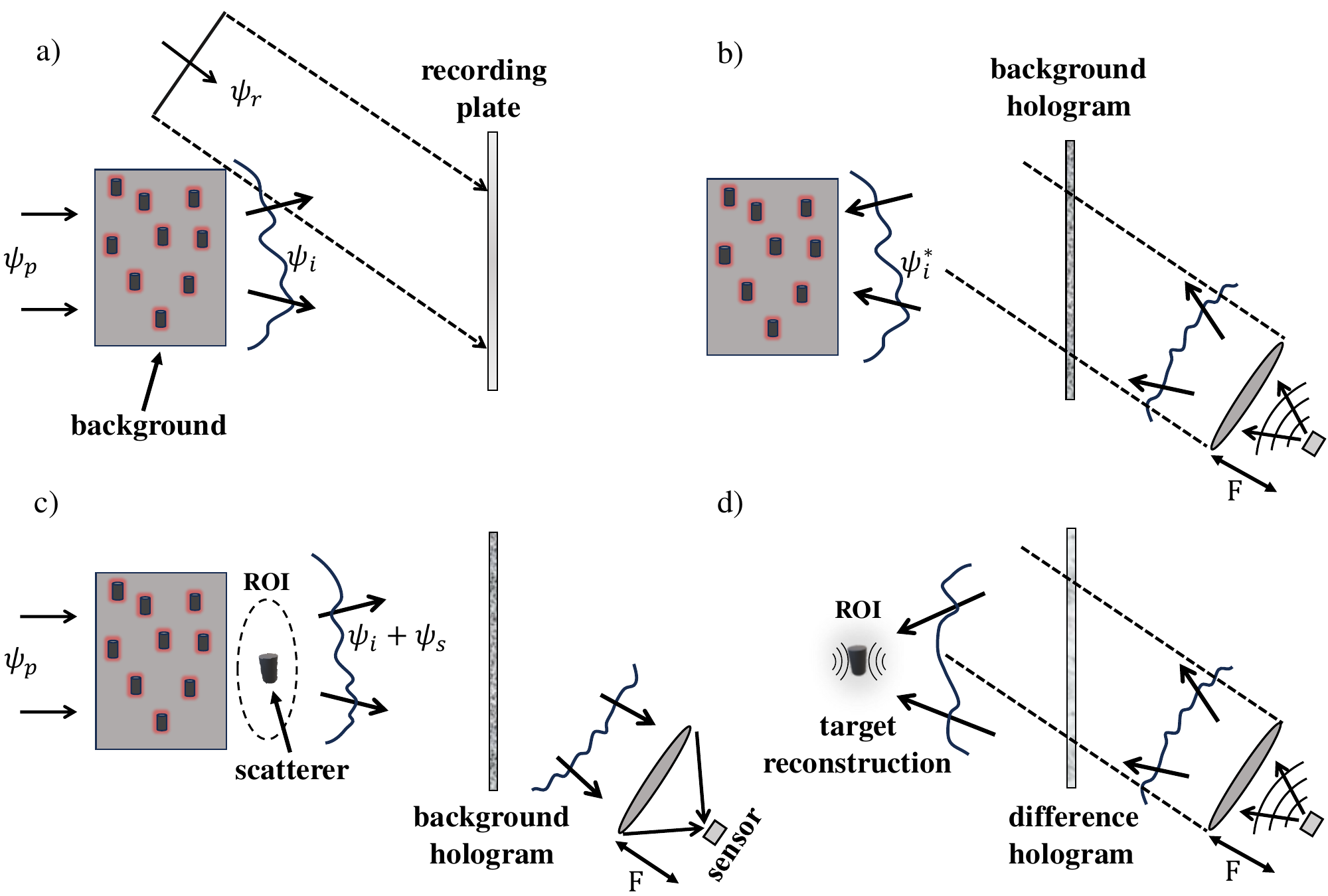}
    \caption{Optical theorem detector based on Leith-Upatnieks-holography.}
    \label{fig_theory}
\end{figure}

\section{Optical-Theorem-Based Holographic Detection}
Consider a complex scattering medium, which may be composed, e.g., of a collection of randomly positioned multiply-interacting scatterers. When this medium is illuminated with probing wave $\psi_p({\bf r})$, where ${\bf r}$ denotes position, then secondary sources are induced in the medium which give rise to scattering. The total field in the medium, which we term ``background field'', is then given by 
\begin{equation}
    \psi_i({\bf r})=\psi_p({\bf r})+\psi_c({\bf r}) \label{eq_jan29_2025_1}\end{equation}
    where $\psi_c({\bf r})$ represents the scattered field due to the background alone. This is illustrated in Fig.~\ref{fig_theory}(a). As shown in Fig.~\ref{fig_theory}(c), if a scattering object or particle appears subsequently in this region, it launches further perturbative fields, henceforth to be denoted as $\psi_s({\bf r})$ (the ``scattered field''), in response to the total illumination $\psi_i({\bf r})$ (the ``incident field''). 
Obviously, the magnitude of the scattered power can be measured from sensing of the scattered fields in all directions. 
On the other hand, more realistically one employs finite sensing apertures only, and thus the practical question is up to what extent it is also experimentally viable to measure or estimate such scattering power from finite-size sensors or through sensing apertures with limited-view access to wave information related to the scene of interest. More generally, ideally the sensing should address the entire extinction, accounting for both scattered power (into all directions) and losses or dissipation (inside the target or particle). The answer to this important question is provided by the optical theorem, which provides the required fundamental principle for the remote sensing, at a realistic limited-view aperture, of the entire extinction. Importantly, the most general forms of the optical theorem (see, e.g., \cite{born2013principles} 
and the references therein) apply to both arbitrary incident fields and quite arbitrary lossless media. 
In particular, it is well known (see, e.g., \cite{marengo2015nonuniqueness} and references therein) that the extinction power $P_e$ is measurable from the scattered field $\psi_s$ in a spatially-limited aperture through sensing mode 
$R({\bf r})$, in particular, 
\begin{equation}
    P_e = - C k_0^{-1} \Im \int d{\bf r} R({\bf r}) \psi_s({\bf r}) \label{eq_jan29_2025_3}\end{equation}
 where $C$ is a constant, $\Im$ denotes the imaginary part, $k_0=\omega/c_0$ is the wavenumber in free space, where $\omega$ is the angular oscillation frequency and $c_0$ is the speed of light, and where $R({\bf r})$ emits (in radiation, acting as a source in the same background medium) the complex conjugated (c.c.) 
 version of the incident field, $\psi_i^*({\bf r})$, in particular, 
\begin{equation}
\int d{\bf r'} R({\bf r'}) G({\bf r},{\bf r'}) = \psi_i^*({\bf r}) \quad {\bf r} \in {\rm ROI}
    \label{eq_jan29_2025_2}
\end{equation} 
where $G({\bf r},{\bf r'})$ is Green's function in the background. 
Thus the key is that this optical theorem receiver ($R({\bf r})$) must be designed such that it is capable of generating in emission the c.c. counterpart $\psi_i^*({\bf r})$ of the incident field $\psi_i({\bf r})$ in the ROI. 
It follows that, for the practical purposes of the sought-after optical realization, the only requirement is the realization of an imaging system wherein the probing source and the intensity-only receiver are strategically arranged such that if the receiver acts as emitter, it can effectively synthesize, in the background, the complex conjugated version of the probing fields in the ROI where targets are expected to appear, for the purposes of detection. Interestingly, this can be achieved via classical lens-based imaging systems in combination with holographic techniques. Moreover, in the proposed methodology the sought-after optical theorem detection can be done using a single detector (e.g., a bucket detector, or single pixel camera), a desirable feature for practical implementations.

Figure~\ref{fig_theory} illustrates the different steps of the required optical-theorem-based sensing. The system is designed to perform two functions: 1) the detection of particles, based on the values of the extincted power $P_e$, where large values over a threshold indicate target presence; and 2) the exploitation of the holographically-captured data for the purposes of developing images that enable particle tracking. Moreover, as shown in the computer simulations, it is also possible to implement certain qualitative forms of tracking via the use of optical theorem data only. The first step is the realization of an optical system capable of producing, in emission, the c.c. version of the probing fields due to the illumination of the complex background. Fortunately, holography provides a practical way to achieve this, particularly in the framework of the so-called Leith-Upatnieks holography \cite{born2013principles} as we explain next. 

As shown in Fig.~\ref{fig_theory}, the 
medium is illuminated with probing wave $\psi_p({\bf r})$, giving rise to the perturbed incident field $\psi_i({\bf r})$ (comprising the probing wave ($\psi_p({\bf r})$) plus the resulting background field ($\psi_c({\bf r})$)) which is measured at the holographic plane with the help of a reference wave $\psi_r({\bf r})$. The field at the hologram plane is equal to $\psi_i+\psi_r$ and consequently the field intensity $I_c$ at point $(x,y)$ in the hologram is given by 
\begin{eqnarray}
I_c (x,y) &=& |\psi_i(x,y)+\psi_r(x,y)|^2 \nonumber \\ 
&=& |\psi_i(x,y)|^2+\psi_i(x,y)\psi_r^*(x,y)+\psi_i^*(x,y)\psi_r(x,y)+|\psi_r(x,y)|^2.\label{eq_jan29_2025_5}
\end{eqnarray}
This enables the recording of a transparency $t(x,y)$ at the hologram plane, which constitutes the recorded hologram, in particular, 
it is proportional to $I_c$: 
\begin{equation}
    t(x,y)= B I_c(x,y) \label{eq_jan29_2025_7}
\end{equation}
where $B$ is a constant. 
When this holographic transparency is excited with the c.c. form of the reference wave, $\psi_r^*({\bf r})$, this gives rise to an induced source at the hologram plane, in particular, 
\begin{equation}
    \rho(x,y)=t(x,y)\psi_r^*(x,y) \label{eq_jan29_2025_9}\end{equation}
    whose generated fields have, according to (\ref{eq_jan29_2025_7}), four components. It is well known that these components can be suitably isolated if the reference wave's angle of arrival relative to the information-carrying beam ($\psi_i({\bf r})$) is properly chosen, e.g., for plane wave reference waves this is achieved via the classical Leith-Upatnieks approach. 
    Now, assuming that the holographic system has been so designed, then it follows that the induced source in (\ref{eq_jan29_2025_9}) generates in the ROI the c.c. field $\psi_i^*({\bf r})$, as desired. This principle is illustrated in part (b) of Fig.~\ref{fig_theory}, where a point source located at the bucket detector acts as emitter, so as to launch in the presence of the focusing lens the c.c. form of the reference wave. Assuming that the reference wave is a plane wave, as shown in the figure, then the required source (for the synthesis of the c.c. reference beam) can be created by placing the point source at the focal plane of the lens, as illustrated in Fig.~\ref{fig_theory}(b). When this c.c. reference wave impinges on the hologram, it gives rise to the emission, into the ROI, of the sought-after c.c. form of the incident field, as required for optical-theorem-based sensing. Thus, Figs.~\ref{fig_theory}(a) and \ref{fig_theory}(b) illustrate the procedure for the required data-driven design of a sensor capable of generating in its complementary radiation or emission counterpart the c.c. form of the incident fields in the ROI. This is done in two steps. First, the incident field is recorded, holographically, at the hologram plane, with the help of a reference wave, e.g., the plane wave shown in Fig.~\ref{fig_theory}(a). The sensor is placed such that when it acts as emitter, in the corresponding lens-based imaging system (where the sensor is placed at the focal plane of the lens, as shown in the figure), it launches the c.c. form of the reference wave. Once this is achieved, then the sensor has the capacity to function as an optical theorem (OT) detector, for the measurement of the extincted power, as we explain next. 

    When a scatterer, such as a particle, appears in the ROI, then this causes a perturbation to the bucket detector's intensity. Like the construction procedure for the OT detector, the measurement stage also involves two steps. First, 
    we measure at the OT bucket detector the intensity of the corresponding field, to be denoted as $I_b$. It is given by 
    \begin{equation}
        I_b ({\bf r}_0)= |\psi_b({\bf r}_0)|^2 \label{eq_jan29_2025_10}\end{equation}
        where ${\bf r}_0$ denotes the detector's position. This intensity corresponds to the background field alone. If a target appears in the region, then the total field $\psi_t({\bf r}_0)$ arriving at the detector changes, and is now given by 
        the sum 
        \begin{equation}
            \psi_t({\bf r}_0) = \psi_b({\bf r}_0)+\Delta\psi({\bf r}_0)\label{eq_jan29_2025_12}
        \end{equation}
where $\Delta\psi({\bf r}_0)$ denotes the perturbation field arising from the presence in the ROI of the target. 
In view of (\ref{eq_jan29_2025_12}), it follows that the intensity at the detector becomes 
\begin{equation}
    I_{b+s}({\bf r}_0)=I_b({\bf r}_0)+2 \Re \left [ \psi_b^*({\bf r}_0)\Delta\psi({\bf r}_0 )\right ] + |\Delta\psi({\bf r}_0)|^2.\label{eq_jan29_2025_14}\end{equation}
    It follows from (\ref{eq_jan29_2025_12}) and (\ref{eq_jan29_2025_14}) that if the perturbation field is relatively weak, relative to the background field, in particular, $|\Delta \psi({\bf r}_0)|<<|\psi_b({\bf r}_0)|$, an expected condition for small targets, then 
    \begin{equation}
       \Delta I= I_{b+s}-I_b\simeq 2\Re \left [ \psi_b^*({\bf r}_0)\Delta\psi({\bf r}_0 )\right ]. \label{eq_jan30_2025_1}
    \end{equation}
    Now, it can be shown (see, e.g., \cite{marengo2016holographic}, eq. 30, and \cite{marengo2015nonuniqueness}, eqs. 32 and 33) that the quantity $\Delta I$ in (\ref{eq_jan30_2025_1}) is proportional to $P_e$ in (\ref{eq_jan29_2025_3}), i.e., $\Delta I$ is, apart from a multiplicative factor, equal to the sought-after extinction power $P_e$, as desired. This completes the derivation of the 
    optical-theorem-based holographic detector of the extinction power. In addition, if the system incorporates a second hologram in the presence of the target, then it becomes possible to also achieve backpropagation-based imaging in this scene, via excitation at the sensor's position and the adoption of a difference hologram, given by the difference of the second and first holograms. Moreover, if the system implements continuous recording of the background hologram, for constant updating, then an additional capability emerges: continuous re-imaging of the scene, which focuses via standard backpropagation into the target region, thereby rendering extra target tracking capabilities. 

\section{Computer Simulations}

We illustrate the application of the developed optical-theorem-based holographic system for the detection of scatterers passing by the ROI which is located near a complex scattering medium. The assumed medium complexity is in general detrimental in a number of tasks based on conventional imaging. However, the same complexity, which generally involves strong multiple scattering or multipath interactions, also brings a high degree of wave interactivity with the target. This feature can, for a suitably designed system, enhance target detectability, as desired. Thus, small cross-section targets that may be hard to detect in free space, or other simple background media, become highly detectable with this system thanks to the embedding, within the system, of the corresponding complex multiple scattering material. Thus the complex medium is integral to the sensing apparatus. Moreover, even though the focus in the following is the situation of an invariant background, the same approach is also applicable for more general varying media, in which case data updates need to be constantly incorporated into the detection process, of course. This concept is implicit in some of the illustrations pertinent to tracking, where it is assumed that the {\em in-situ} background holograms are captured continuously. 

\begin{figure}[H]
    \centering
    \includegraphics[width=1\linewidth]{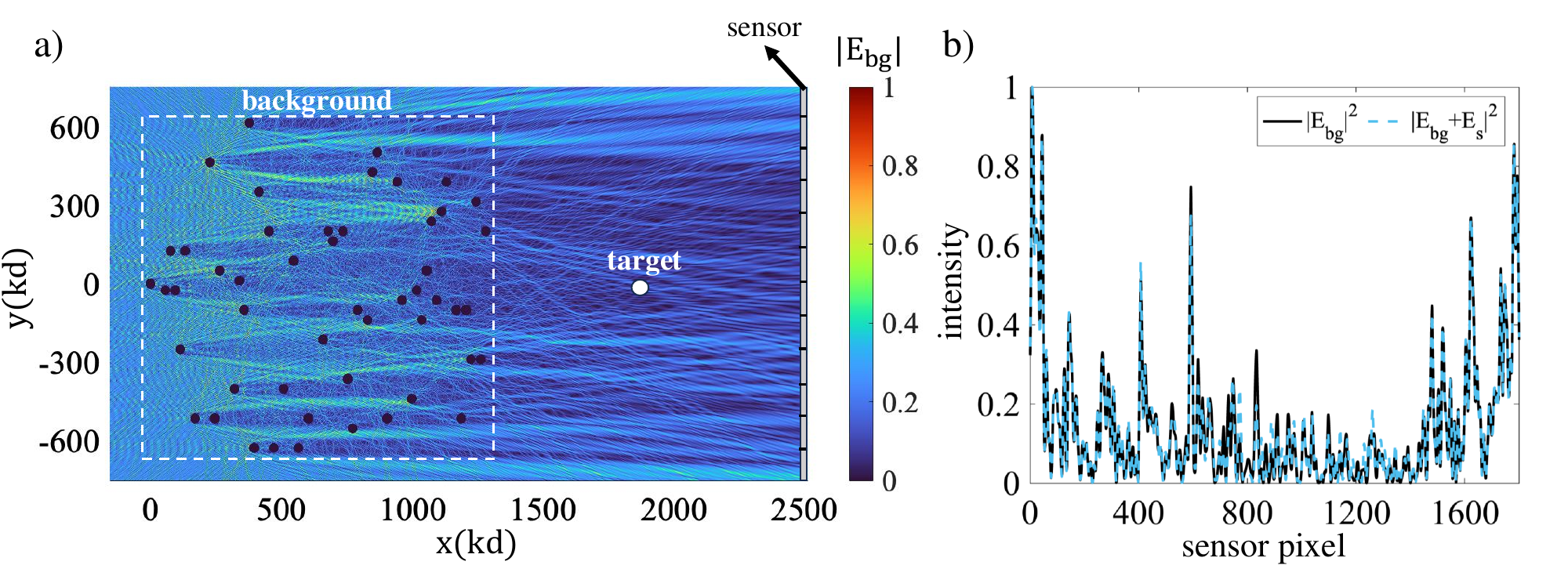}
    \caption{a) Background medium and the corresponding field intensity. b) Background field intensity and total (background plus scattered field) intensity at the hologram aperture. }
    \label{detection_alone}
\end{figure}
In the simulations, electromagnetic scattering is modeled in two-dimensional (2D) space, for a complex medium consisting of a collection of randomly distributed, parallel dielectric cylinders, whose multiple scattering interactions are fully incorporated via standard methods (see, e.g., \cite{tsuei1988multiple}, \cite{olaofe1970scattering}, and the references therein). Of particular interest is the detection of a target or ``particle'' crossing the vicinity of the complex medium. The problem of detecting two or more targets (e.g., a stream of particles) is also examined, and the unique response characteristics of different random sets of cylinders are also briefly discussed. 
\begin{figure}[H]
    \centering
    \includegraphics[width=1\linewidth]{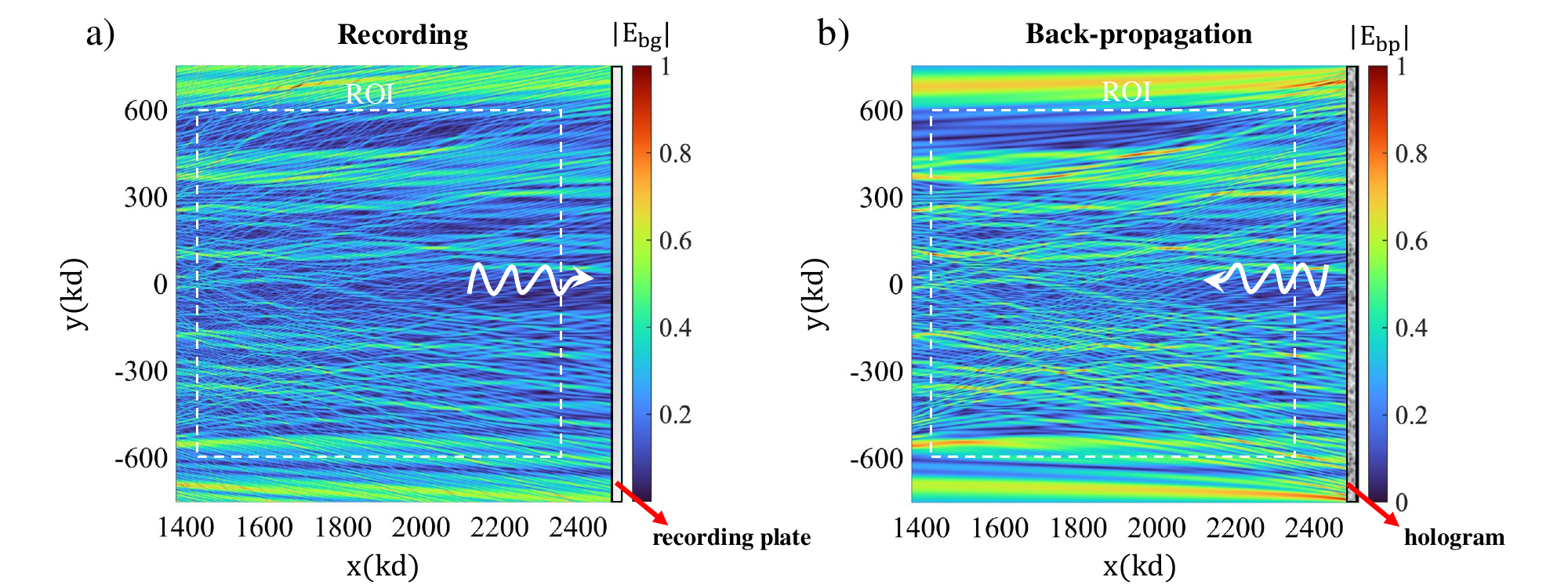}
    \caption{ (a) Recording stage of the hologram generation. (b) Background field reconstruction in the ROI when the bucket detector acts as a source.}
    \label{tr}
\end{figure}
Electromagnetic scattering involving multiple objects is quite challenging due to the need to account for coupling between the elements. To satisfy the relevant boundary conditions on the surface of the cylinders we made use of the well-known translational addition theorem. The latter allows the representation of the scattered fields from one cylinder as incident fields on another one, facilitating the use of a common reference origin for both. There is extensive literature and numerous studies focused on calculating scattering from cylindrical scatterers, as well as on analyzing multiple scattering interactions within groups of such scatterers \cite{twersky1952multiple,zitron1961higher}. A convenient way to model this problem is to use transverse magnetic (TM) and transverse electric (TE) scalar potentials that can be calculated analytically for a large number of scatterers \cite{olaofe1970scattering, hulst1981light}. In particular, we considered a TM incident wave ($\text{H}_\text{z} = 0$) of the form 
\begin{align}
    u^{i}_{j} = -\frac{1}{k} \sum_{n=-\infty}^{\infty} i^{n+1}[J_{n}(\rho_j)\epsilon_{j}] \exp{(-in\gamma_{j})}
\end{align}
where $(r_j,\gamma_j)$ are the polar coordinates with respect to the origin of $j$-th scatterer, $\rho_j = k r_j$, and $J_n(\cdot)$ is the Bessel function of order $n$, and $\epsilon_{j} = \exp{(id_{1j}k\cos{(\beta_{1j})})}$ is the phase-shift factor for the incident wave where $d_{1j}$ is the separation between the first cylinder and the $j-$th one. Also, $\beta_{1j}$ is the angle that the line connecting the centers of the first and $j-$th cylinders makes with the x-axis. To continue, one needs to write the electromagnetic boundary conditions for every one of the scatterers. This gives rise to suitable continuity conditions that $u$ and the normal derivative $ \partial u/\partial r$ must obey. As mentioned, the translation addition theorem facilitates this process. The scattered fields in the exterior part of the cylinders can be written as
\begin{equation}
    u^{s}_{j} = \frac{1}{k} \sum_{n=-\infty} ^{\infty} i^{n+1} [H_{n}(\rho_{j})] \exp{(-in \gamma_{j})} {}_{j}{b_n}
\end{equation}
where ${}_{j}{b_n}$ are the corrected scattering coefficients which are obtained from the boundary condition at $r_j = a_j$. Also, $H_n(\cdot)$ is the Hankel function. It should be noted that in the absence of multiple scattering (i.e., for a single cylinder) the single-scattering coefficients are already known, of course \cite{devaney2012mathematical}.
\begin{figure}[H]
    \centering
    \includegraphics[width=1\linewidth]{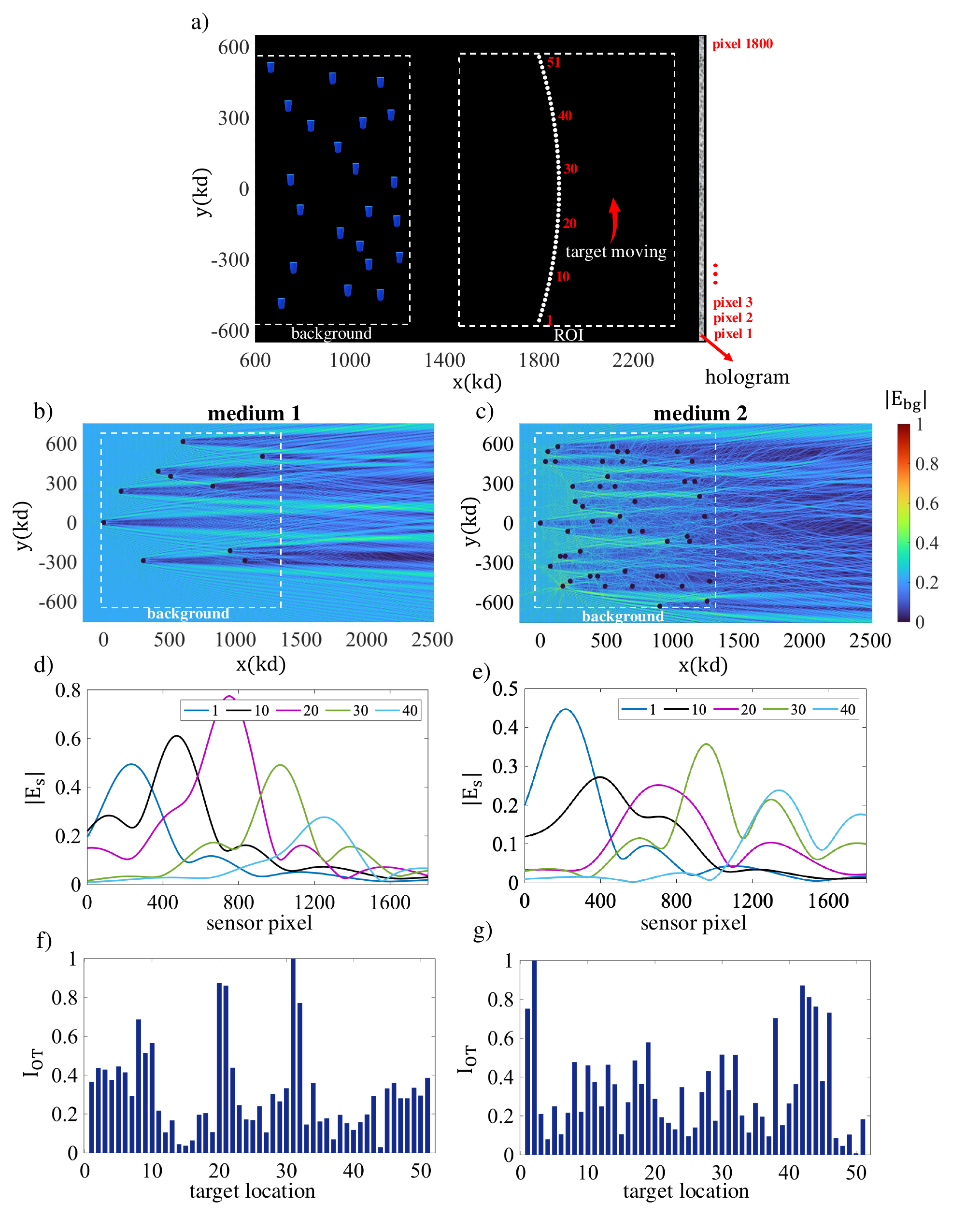}
    \caption{ Intrusion scenario where a single cylindrical particle enters the ROI and traverses a radial path. (a) Two random background media with different scatterer densities are utilized in this demonstration which are called medium 1 (eleven scatterers) and 2 (51 scatterers). (b,c) Normalized intensity of the total field that the background media generate throughout the ROI. (d,e) Intensity of the scattered fields from the moving target present in selected locations on the hologram aperture. (f,g) Distribution of the normalized value of the OT detector when medium 1 and 2 are used, respectively.}
    \label{single_target} 
\end{figure}
Furthermore, the TM potential in the immediate vicinity of cylinder $C_{j}$ consists of three components including primary and secondary incident, and scattered waves. If proper boundary conditions are applied to these equations then the corrected scattering coefficients can be computed as
\begin{align}
    {}_{l}{b_n} = {}_{l}{b_n}^{(0)}[\epsilon_{l} + i^{n+1} \sum_{\substack{l=1 \\ l \neq j}}^{N}\exp{(in\beta_{lj})_{jl} B_{-n}}] \nonumber \\ 
    {}_{lj}{B_n} = \sum_{s=-\infty}^{\infty} i^{s+1}\Psi_{lj}\exp{(-is\beta_{lj})} H_{n+s}(\delta_{lj}){}_{l}{b_{s}} 
    \label{coupled}
\end{align}
Where ${B_n}$ is the coupling term and $\Psi_{lj}=(-1)^s$ if $j>l$ and $(-1)^n$ otherwise. These two equations are coupled and can be solved either iteratively or using inverse methods depending on the separation between the cylinders. The most accurate solution is obtained through the inverse method where the mentioned coupled equations should be written in a matrix form
\begin{equation}
    M = (I-C)^{-1} S
\end{equation}
Where $S$ and $M$ denote vectors of single and multiple scattering coefficients. $C$ is the coupling matrix that contains the coupling terms between single and corrected  scattering coefficients in equations \ref{coupled}. By solving this matrix inverse equation, the corrected scattering coefficients are obtained which completes the parameters required for finding the scattered fields in the simulation region. 

Figure~\ref{detection_alone}(a) shows the geometry considered in the simulations. It illustrates the background medium which is 
composed of a cluster of cylinders with random positions and refractive indices $3<\text{n}<4$ with radius of $\text{r}_\text{s} = 3\lambda$. The scatterers are placed in the highlighted rectangular region which has dimensions $0<l_x<400\pi \lambda$ and $-200\pi \lambda<l_y<200\pi \lambda$. The ROI is assumed to be to the right of the scattering region, as shown in the figure. As shown in the figure, the intensity of the corresponding background field in that region is quite complex, as expected from the adopted medium's complexity. Part (b) of the figure shows the field intensity profile at the hologram aperture, for both the background field and the total (background plus scattered field). 
The target consists of a single cylinder with refractive index and radius of $\text{n}=1.2$ and $\text{r}_\text{s}=1.5\lambda$, respectively. 
The two plots are very similar, for the target considered. This shows that, at least for weak scattering targets, it is very difficult to detect the target presence based on intensity-only measurements at the hologram plane.   
In contrast, the optical theorem method adopted in this work is a coherent detection method and, as we show next, it does reveal both the target presence as well as unique features that can be associated with the target's trajectory.

\begin{figure}[H]
    \centering
    \includegraphics[width=1\linewidth]{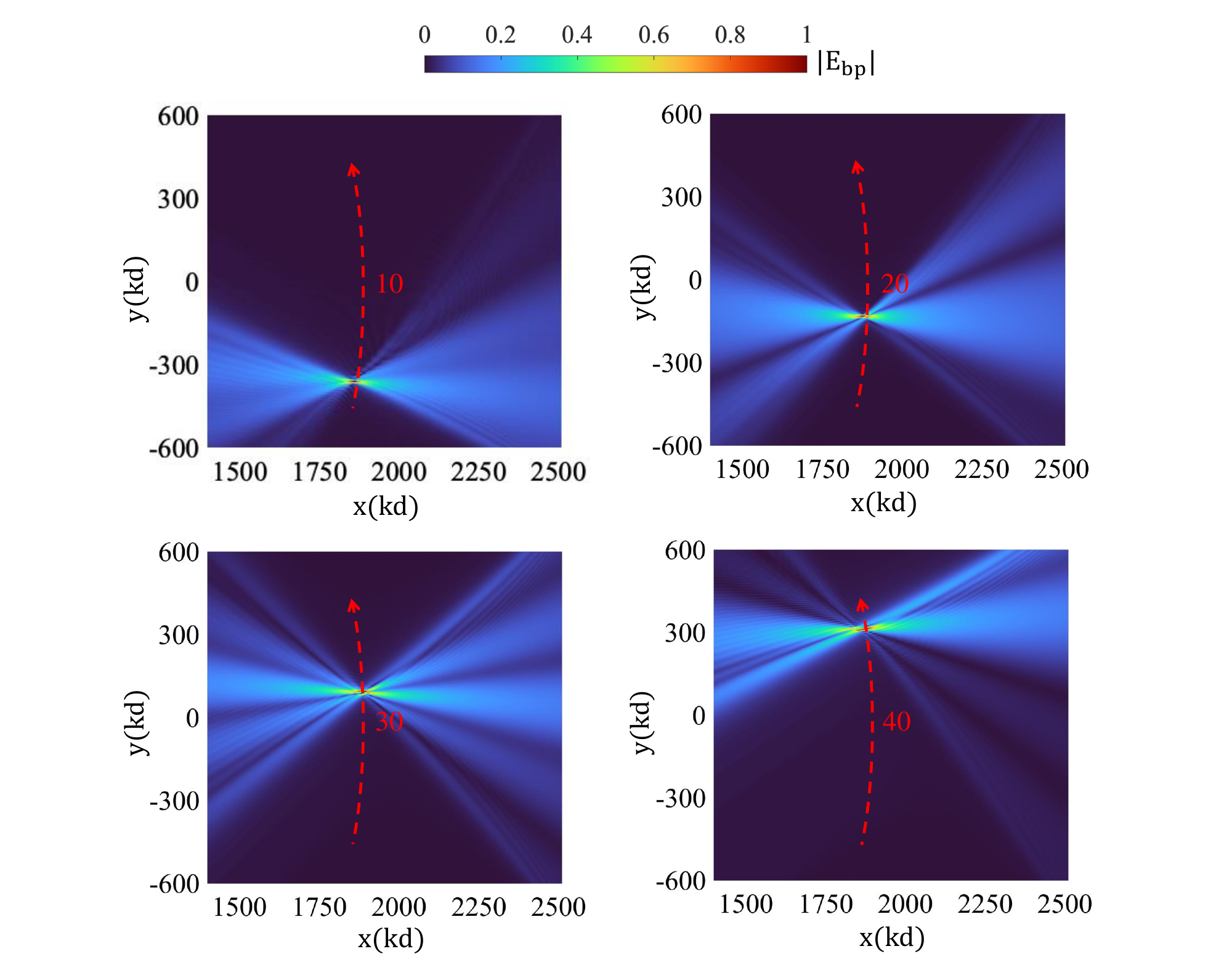}
    \caption{Tracking of the particle via backpropagation-based imaging based on the difference hologram.}
    \label{fig_tracking}
\end{figure}
The effectiveness of the optical theorem approach as a method to detect the scattering cross section rests on the ability of the sensor system to radiate, in the complementary transmission mode, the c.c. version of the probing field adopted to interrogate the target. The key attribute is the system's ability to reproduce, in transmission, the c.c. field at least within the ROI where targets are expected to appear. Figure~\ref{tr} demonstrates the viability of the system in Fig.~\ref{detection_alone} for the effective synthesis of the c.c. probing field in the desired ROI. Figure \ref{tr}(a) shows a plot of the intensity of the incident wave probing the ROI. The figure shows the hologram plane, where the field is recorded holographically with the help of a reference wave. Part (b) of the figure shows the corresponding backpropagation reconstruction, corresponding to the synthesis of the c.c. field using the same hologram recorded in part (a), upon excitation with the relevant c.c. form of the reference wave. This is achieved by launching a probing field from the single-pixel or bucket detector region, which upon interacting with the lens gives rise to the c.c. form of the reference wave adopted in the hologram generation step. Comparing the two plots of the field intensity, we conclude that they are very similar, as desired. Thus, the required c.c. field reconstruction is successful with this sensing apparatus, as required for applicability of the optical theorem sensing approach.

Another feature of the developed method is that it also allows inferences about the target location, its trajectory, as it moves in the region. As an example, we have considered a scenario where the cylindrical particle ($\text{n} = 1.2$, $\text{r}_\text{s} = 1.5\lambda$) moves in a radial path ($\text{r}=600\pi \lambda$, $ -0.3< \theta < 0.3$ (radians)). The simulation geometry is illustrated in Fig.~\ref{single_target}(a). We examined the tracking capabilities of the system with different methods. First we discuss the role of medium complexity, in both detection and tracking, using the OT approach only. We compared, for this purpose, two media, with different numbers of scatterers. Parts (b) and (c) of Fig.~\ref{single_target} depict the configurations of 1) background medium 1 which contains ten scatterers and 2) background medium 2 which contains fifty scatterers. In these experiments, we considered refractive indices in the  range $3 < n < 4$. The positions were selected randomly to simulate complex random media. As can be seen in these figures, the medium with more scatterers generates a more complex probing field in the ROI, as anticipated. In addition, Figs.~\ref{single_target}(d,e) show the scattered fields corresponding to different target positions. Given the size and shape of the scatterer adopted in the experiments, it is expected that it should have a relatively symmetric scattering pattern in the forward direction. This is indeed consistent with the results in Figs.~\ref{single_target}(d,e). The OT results are shown in Figs.~\ref{single_target}(f,g). The OT indicator is normalized in these plots, to enhance interpretation. 
The pattern of computed OT indicator in the case of medium 2 clearly shows more abrupt variations as a function of target position. Thus increased medium complexity translates to greater variability in the measured OT response. This could be useful in detecting nanoscale movement of the particles, as even a tiny variation in position gives rise to a noticeable change in the OT indicator. In addition, if the adopted medium's OT response is stored, this could be used as a template for estimation of the target position, e.g., the peaks in the OT plots can be associated to the maxima of the stored background field, to detect proximity of the target to given reference points, as desired. We discuss next another tracking technique that relies of difference holograms, based on two background holograms captured at different times. These can be consecutive times, or we can use the background hologram of a prior past time as the reference, for the image generation process. 

\begin{figure}
    \centering
    \includegraphics[width=1\linewidth]{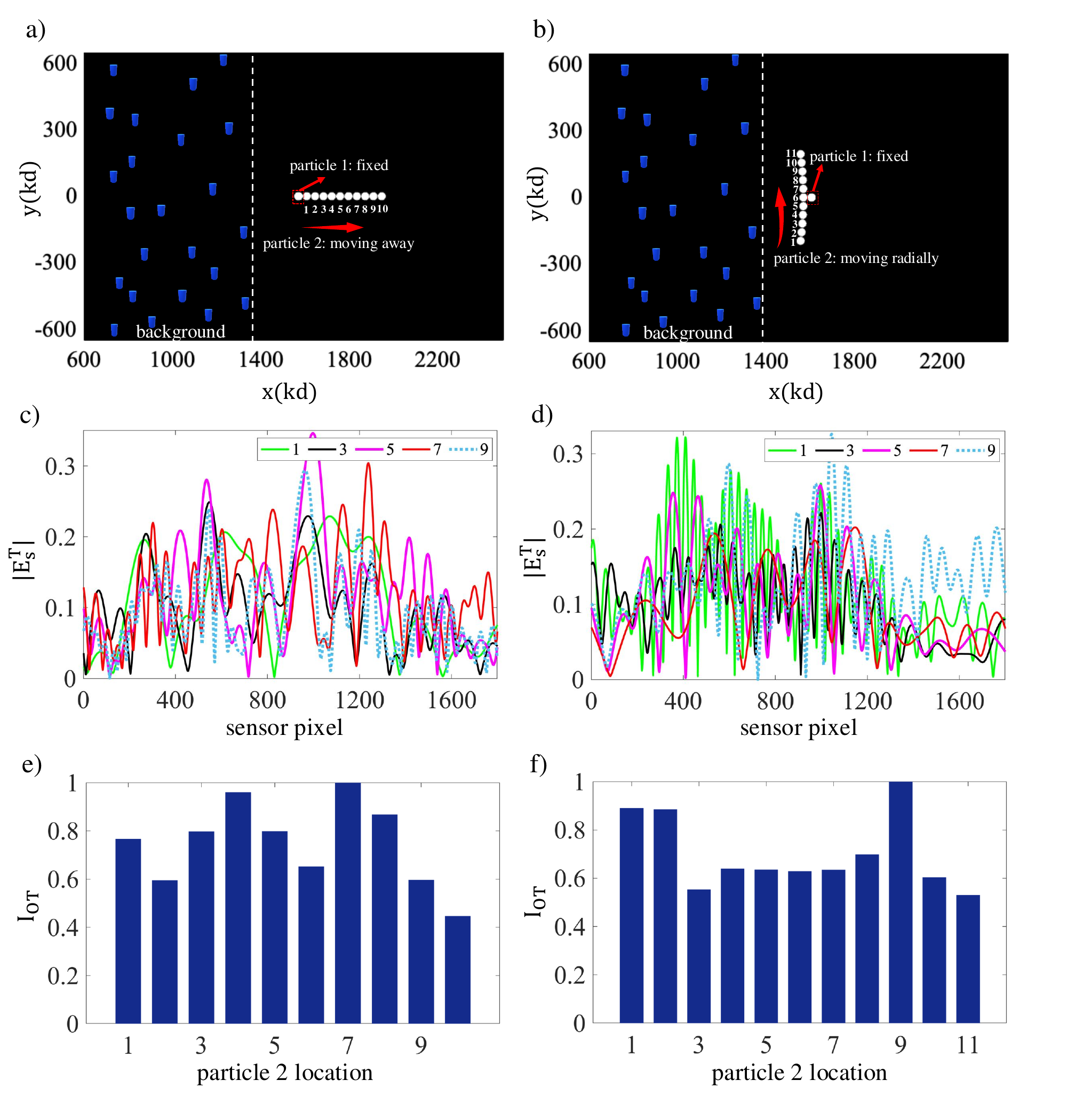}
    \caption{ Analyzing the effect of relative movements of two already present cylindrical particles on the observed OT value. (a,b) Configuration adopted in the experiments. One of the particles is kept at a fixed location. The second particle moves in the trajectory shown. (c,d) Intensity of the total scattered field from both particles in selected locations of the moving particle. As can be seen this total scattered field $\text{E}^{\text{T}}_{\text{s}}$ is highly dependent on the relative positions of the particles mainly due to multiple scattering effects between them. (e,f)  OT indicator versus target position.}
    \label{dual_target}
\end{figure}

Figure~\ref{fig_tracking} shows the backpropagation-based imaging based on the difference hologram corresponding to two different times. This can be, e.g., the reference background hologram corresponding to the medium, plus an additional hologram corresponding to the data-gathering steps, where one can measure not only the OT sensor intensity but also capture a second hologram pertinent to the scene at the same time interval. The latter hologram includes the effect of both the background and the target, and therefore the subtraction of the two holograms, which we term ``difference hologram'' contains information about the target whereabouts. Under point-source-like excitation at the bucket detector, it is possible to generate a real-image reconstruction of the target in the ROI, and this can be done, e.g., with a tandem holographic system, implementing the complementary function of tracking. Alternatively, this can be done computationally from said difference hologram, via conventional backpropagation from the hologram aperture. The backpropagation images shown in the figure correspond to the corresponding field intensity and are quite impressive. They consistently focus on the correct target position, as it trespasses the region along the trajectory shown. 

In the last simulation of this paper (Fig.~\ref{dual_target}) we considered a scenario that further illustrates additional features of the entire approach. In this case we considered two targets, in two different trajectories: horizontal motion in part (a) of the figure; and vertical movement in the companion part (b). One of the particles remains fixed, while the other particle moves in the shown trajectory. As the plots reveal, the scattered field changes a lot from one sense of motion to the other. The OT indicator also exhibits noticeable differences in these two trajectories. Motivation was provided by a possible ``target separation problem'' where a target may appear, effectively, as a single entity initially, as depicted in part (a) of the figure, but become subsequently separated into two different targets. As the figure reveals, the OT responses of the two possible trajectories are quite distinguishable, and clearly they can, in principle, be discerned via the OT indicator if prior stored information is available. In addition, we also ran additional simulations (results not shown) of the associated backpropagation imaging (as in Fig.~\ref{fig_tracking}), and we found that the targets become clearly differentiated in the images, as desired.

\section{Conclusion}

In this work we have developed an optical-theorem-inspired holographic method for the detection and tracking of particles embedded in complex multiple scattering media. Motivation is provided by a number of applications, e.g., in the biomedical, environmental, and defense areas, where one seeks to detect and characterize targets of interest, such as particles, aerosols, etc., in real-time and under evolving conditions, and where the embedding medium is typically complex and random. In this work we have adopted the extinction cross section as the key physical descriptor of the particle's presence, under the assumption that the sought-after minimally detectable target size or associated extinction is approximately known. This permits, in principle, the suitable calibration of the sensing apparatus, through the proper threshold level selection, so as to render the desired probability of detection and false alarm under the anticipated conditions, e.g., trade-offs, risk levels, etc. The classical optical theorem of scattering theory provides the underlying physical principle for the proposed detection system. In this work we proposed the use of a holographic system, where the reference wave is an oblique-incidence plane wave arriving at the hologram plane. The well-known principles of Leith-Upatnieks holography are therefore applicable, so as to enable the critical isolation of the four different components of the field when the hologram is illuminated either by the reference wave or the c.c. form of that wave, for the purposes of image reconstruction. The detector itself is formed by a single-pixel camera or bucket detector, positioned at the focal plane of the lens, for the creation of a lens-law-obeying imaging system with the laser source that generates the reference wave. This ensures that the bucket detector emits, in radiation, the c.c. form of the reference wave, as required for applicability of the optical theorem. This enables, in turn, the launching of the c.c. form of the background medium field, through illumination of the ``background medium hologram'' with this c.c. reference wave, thereby completing the requirements that the sensor must obey in order to function as a detector of the scattering extinction, as desired. 

The proposed detection and tracking methods have been illustrated numerically, incorporating all the pertinent multiple scattering effects, for background media composed of random collections of dielectric nanocylinders. 
We have shown that the OT approach enables the sought-after particle detection under conditions where target presence is virtually invisible in the associated intensity-only image at the hologram plane. Moreover, although not shown in the paper, this signal enhancing effect associated to the OT method is even more noticeable under noisy conditions, since the OT approach is based on coherent processing in which noise cancellation is very effective, as is well known. We  have also studied the dependence on target position of the calculated extincted power as well as the effects associated to movement direction and the role of additional targets in the region. The results indicate that the OT approach is also potentially applicable for the associated task of tracking. In addition, the obtained results also illustrated the use of the difference hologram, corresponding to successive hologram captures, for the backscattering-based imaging of the target, through which tracking can also be achieved. This can be done either computationally (synthetically) or through a companion, in tandem apparatus, for real-image reconstruction based on said stored holograms. 

The derived detector is single-pixel, and it also relies on exploitation of the background medium itself. In that sense, it can be thought of as an {\em in-situ} compressive detector: it measures the extinction, a physical indicator of target presence, via a single, well-localized sensor, that is embedded into, or is part of the surrounding medium of interest. The data-driven nature of this sensing approach also makes it universally applicable, and indeed the sensor can be thought of as being adaptive to the medium. Moreover, the adaptability properties of the derived optical-theorem-based system also make it viable for real-time detection and tracking in dynamic, varying environments. In particular, in the derived approach, the incident field, associated to the background, is recorded holographically, and this operation can be implemented constantly, at regular intervals, for continuous updating of the associated optical-theorem-holograms through which measurement of the extinction becomes possible at the single-pixel sensor. This possible dynamicity of the hologram itself is also what enables the nonstop updating of the associated ``difference hologram'' through which imaging of the target becomes accessible, despite the medium's complexity and dynamic variability. Future directions include the possible application of the results derived in this paper in other fields such as physical layer security, encryption in complex media, secure optical-theorem-based communications, and related areas. We are currently exploring these areas, and plan to report on the associated research developments in the future. 
 
\bibliography{main}
\bibliographystyle{unsrt}
\end{document}